\documentclass[runningheads]{llncs}
\usepackage[latin1]{inputenc}
\usepackage[T1]{fontenc}
\usepackage{stmaryrd,pifont,yhmath,wasysym}
\usepackage{amsmath,amstext,amssymb,amsfonts}
\usepackage{graphicx,xcolor,multirow, fancybox,verbatim,xspace,}
\usepackage{tikz,pgflibraryarrows,pgffor,pgflibrarysnakes}
\usetikzlibrary{snakes}
\usepgflibrary{shapes}
\usepackage{subfigure,enumerate}
\usepackage{bussproofs}
\usepackage{macros}
\usepackage{hyperref,url}
\usepackage{enumitem}
\begin{document}
\title{Zone-based verification of timed automata: extrapolations,
  simulations and what next? \thanks{This work was partially funded by ANR project Ticktac
(ANR-18-CE40-0015).}}
\titlerunning{Zone-based verification of timed automata}
\author{Patricia Bouyer\inst{1} \and Paul
  Gastin\inst{1,5} \and Fr\'ed\'eric
  Herbreteau\inst{2} \and \\ Ocan
  Sankur\inst{3} \and
  B. Srivathsan\inst{4,5}}
\authorrunning{P. Bouyer, P. Gastin, F. Herbreteau, O. Sankur, B. Srivathsan}
\institute{Universit\'e Paris-Saclay, CNRS, ENS Paris-Saclay,
  Laboratoire M\'ethodes Formelles, Gif-sur-Yvette, 91190, France 
  \and Univ. Bordeaux, CNRS, Bordeaux INP, LaBRI, UMR5800, Talence, 33400, France
  \and Univ Rennes, Inria, CNRS, Irisa, France 
  \and Chennai Mathematical Institute, India 
  \and CNRS IRL 2000, ReLaX, Chennai, India}
\maketitle              
\begin{abstract}
  Timed automata have been introduced by Rajeev Alur and David Dill in
  the early 90's. In the last decades, timed automata have become the
  \textit{de facto} model for the verification of real-time
  systems. Algorithms for timed automata are based on the traversal of
  their state-space using zones as a symbolic representation. Since
  the state-space is infinite, termination relies on finite
  abstractions that yield a finite representation of the reachable
  states.

  The first solution to get finite abstractions was based on
  extrapolations of zones, and has been implemented in the
  industry-strength tool \uppaal{}. A different approach based on
  simulations between zones has emerged in the last ten years, and has
  been implemented in the fully open source tool \tchecker{}. The
  simulation-based approach has led to new efficient algorithms for
  reachability and liveness in timed automata, and has also been
  extended to richer models like weighted timed automata, and timed
  automata with diagonal constraints and updates.

  In this article, we survey the extrapolation and simulation
  techniques, and discuss some open challenges for the future.

\end{abstract}

\section{Introduction}\label{sec:introduction}

Timed automata have been defined in the early 1990's~\cite{AD90,AD94}
as a formal model for representing systems in which timing constraints
play an important role. A timed automaton is a finite automaton that
can manipulate variables. These variables have very specific
behaviours: they increase synchronously with time, they can be
constrained (for instance, compared to a constant), and they can be
updated (most often, reset to $0$).  The number of states of a timed
automaton is infinite, and hence no properties of standard finite
automata can trivially be transferred to timed automata. However, a
finite-state abstraction, called the \emph{region automaton}, can be
constructed that allows us to check for reachability properties,
$\omega$-regular properties, some branching-time timed temporal
logics, \textit{etc}.  An extensive literature has been written on
this model since the original paper~\cite{AD90}.  Research on timed
automata since then has resulted in a rich theory, and in more
practice-guided developments. In this paper we focus on the latter,
and present algorithmic issues, which we believe are the important
ones.

There are two main families of symbolic approaches to analyze
reachability properties: the backward and the forward analysis.
Backward analysis consists in computing iteratively (symbolic)
predecessors of the target state, until hitting the initial
configuration or until no new symbolic state can be computed. By
chance, such a symbolic computation, when applied to timed automata,
always terminates and therefore provides a symbolic algorithm for the
verification of reachability properties. Unfortunately it is not
really adequate to analyse models with discrete structures (like
discrete variables) and does not extend very well to other properties.
On the other hand, forward analysis consists in computing iteratively
(symbolic) successors of the initial configuration, until hitting the
target or until some stopping criterion. The exact forward computation
does not terminate in general, and so there is a need for some
theoretical developments to ensure termination of the computation and
prove the correctness of the approach. We focus on the forward
analysis algorithm here, and primarily restrict to reachability
properties.

The key advantage in timed automata verification is the fact that the
reachable symbolic states can be represented using simple constraints
over the variables, called zones. Zones can be efficiently
manipulated, and therefore the computation of successors is quick. The
main challenge is to get a stopping criterion for the forward analysis
that can ensure, as soon as possible, that all reachable states have
been visited. The first approach to stop the forward analysis makes
use of an extrapolation operation on zones.  Extrapolation of a zone
gives a bigger zone picked from a finite range, thereby automatically
ensuring termination. Starting from the late 90's, extrapolations have
been studied in depth, leading to the development of efficient
tools~\cite{BDL+06,BDM+98} and several fundamental observations about
what extrapolations can do and cannot
do~\cite{bouyer03,bouyer04,BBLP04,BBLP05}. With the aim of
circumventing the limitations of the extrapolation-based approach, an
alternative solution has been pioneered in~\cite{HSW12,HSW16}. This
approach keeps the zones as they are, and makes use of simulations
between them to stop the computation. In addition to overcoming the
limitations of the extrapolation-based approach, the simulation-based
approach has paved the way for several remarkable advances in timed
automata verification, in particular: dynamic abstractions for
reachability checking~\cite{HSW13,RSM-cav19}, refinement-based
verification of liveness properties~\cite{HSW16}, simulation-based
finite abstractions for weighted timed automata~\cite{BCM16}, for
automata with diagonal constraints and more general updates than plain
resets \cite{GMS18,GMS19,GMS20}, for pushdown timed
automata~\cite{AGP21} and for event-clock automata \cite{AGGS22}. Some
of these techniques have been implemented in the tool
\tchecker{}~\cite{TChecker}, and experiments have shown significant
gains with respect to previous approaches.

The goal of this survey is to present the extrapolation-based and the
simulation-based approaches, which have been instrumental in the
success of timed automata verification. We also include a discussion
on ``what next?'', where we throw light on some of the current
challenges in timed automata verification that cannot be tackled just
by using extrapolations or simulations.  The plan of the paper is the
following. In Section~\ref{sec:defs}, we briefly recall the model of
timed automata and its semantics. In Section~\ref{sec:symbolic}, we
explain the symbolic approach to the forward analysis of timed
automata, by presenting zones, generic forward schemes to enforce
termination of the forward computation while preserving completeness
and soundness, and finally DBMs as a useful representation of
zones. In Section~\ref{sec:extra}, we present the extrapolation-based
approach.  In Section~\ref{sec:simulation}, we present the
simulation-based approach.  In Section~\ref{sec:more}, we discuss two
extensions, one concerning the model, and one concerning the
properties to be checked.  There have been multiple tools which have
been developed to verify timed automata; in Section~\ref{sec:tools} we
present \uppaal, the most successful model-checker, whose development
started in 1995~\cite{BLL+95} and is based algorithmically on the
extrapolation approach, and \tchecker, an open-source model-checker
under development~\cite{TChecker}, which implements the most recent
algorithms based on simulation. In Section~\ref{sec:next} we discuss
the next challenges which occur as a natural follow-up to the theory
and practice that has been built so far.

\section{Birth of timed automata for verification: the early 90's}
\label{sec:defs}

\subsection{Preliminaries}

We consider a finite set $X$ of variables, called \emph{clocks}. A
\emph{(clock) valuation} over $X$ is a mapping
$v\colon X \rightarrow \IR_+$ which assigns to each clock a
non-negative real, which denotes a time value. The set of all clock
valuations over $X$ is denoted $\IR_+^X$, and $\mathbf{0}$ denotes the
valuation assigning $0$ to every clock $x \in X$.
Let $v \in \IR_+^X$ be a valuation and $\delta \in \IR_+$, the
valuation $v+\delta$ is defined by $(v+\delta)(x)= v(x)+\delta$ for
every $x\in X$. For $Y \subseteq X$, we denote by $[Y]v$ the valuation
such that for every $x \in Y$, $([Y]v)(x) = 0$ and for every
$x \in X \setminus Y$, $([Y]v)(x) = v(x)$.

Given a finite set of clocks $X$, we introduce two sets of \emph{clock
  constraints over $X$}. The most general one, denoted $\calC(X)$, is
defined by the grammar:
\begin{eqnarray*}
  g & \ ::= \ & x \bowtie c\ \mid\ x-y \bowtie c\ \mid\ g \wedge g\
  \mid\ \textit{true} \\ & & \text{where}\ x, y \in X,\ c \in \IZ\
  \text{and}\ {\bowtie} \in \{<,\leq,=,\geq,>\}.
\end{eqnarray*}
A clock constraint of the form $x-y \bowtie c$ is said to be a
\emph{diagonal} constraint. Next we also use the proper subset of
\emph{diagonal-free} clock constraints where diagonal constraints are
not allowed. This set is denoted $\calCdf(X)$.

If $v \in \IR_+^X$ is a clock valuation, we write $v \models g$ when
$v$ satisfies the clock constraint $g$, and we say that $v$ satisfies
$x \bowtie c$ (resp.\ $x-y \bowtie c$) whenever $v(x) \bowtie c$
(resp.\ $v(x)-v(y) \bowtie c$). If $g$ is a clock constraint, we write
$\val{g}$ for the set of clock valuations
$\{v \in \IR_+^X \mid v \models g\}$.

\subsection{The timed automaton model~\cite{AD90,AD94}}

A \emph{timed automaton} is a tuple $\calA = (Q,X,q_0,T,F)$ where $Q$
is a finite set of states, $X$ is a finite set of clocks, $q_0 \in Q$
is the initial state,
$T \subseteq Q \times \calC(X) \times 2^X \times Q$ is a finite set of
transitions, and $F \subseteq Q$ is a set of final states. A timed
automaton is said to be diagonal free if $\calC(X)$ is replaced by
$\calCdf(X)$ in the definition of the transition relation.

A \emph{configuration} of a timed automaton $\calA$ is a pair $(q, v)$
where $q \in Q$ is a state of the automaton and $v \in \IR_+^X$ is a
valuation. The semantics of $\calA$ is given as a transition system
over its configurations. The initial node is $(q_0, \textbf{0})$
consisting of the initial state $q_0$ and the initial valuation
$\textbf{0}$. There are two kinds of transitions:
\begin{description}
\item[time elapse] $(q, v) \xra{\delta} (q, v + \delta)$ for every
  $\delta \in \IR_+$,
\item[discrete transition] $(q,v) \xra{t} (q_1, v_1)$ if there exists
  a transition $t = (q, g, R, q_1) \in T$ such that $v \models g$ and
  $v_1 = [R]v$.
\end{description}

A run of $\calA$ is a (finite or infinite) alternating sequence of
time elapses and discrete transitions starting from the initial
configuration:
$(q_0, \textbf{0}) \xra{\delta_1} (q_0,\textbf{0}+\delta_1) \xra{t_1}
(q_1, v_1) \xra{\delta_2} (q_1,v_1+\delta_2) \xra{t_2} \ldots$.

In the following we will be interested in the verification of
reachability properties: given a timed automaton $\calA$ and a control
state $q$, does there exist a (finite) run that leads to a
configuration of the form $(q, v)$? This problem was shown to be
PSPACE-complete in the paper introducing timed automata~\cite{AD90}.
The first algorithm for reachability made use of a finite partition of
the set of valuations into \emph{regions}, and then using it to
construct a region automaton.  There are exponentially many regions
and hence this solution is not useful in practice.


\section{A symbolic approach to the verification of timed automata}
\label{sec:symbolic}

The original decidability result which uses regions is not suited for
implementation, and symbolic approaches exploring the set of reachable
states (forward or backward), and based on so-called
\emph{zones}~\cite{ACD+92,HNSY94} are preferred. In this survey we
focus on forward computations, which compute iteratively successors of
the initial configuration, until hitting the target set (a
configuration whose state is in $F$), or (hopefully) until getting a
certificate (e.g., an invariant) showing that it cannot be hit.

We fix for this section a timed automaton $\calA = (Q,X,q_0,T,F)$.

\subsection{Zones and symbolic forward computation}

\begin{figure}[h]
  \subfigure[Zone $Z$\label{subfig:Z}]{
    \begin{tikzpicture}[scale=.35]
      \everymath{\scriptstyle}
      \draw [latex'-latex'] (0,6) -- (0,0) -- (10,0);
      \draw [fill=black!20!white] (3,0) -- (4,0) -- (9,5) -- (3,5) -- cycle;
      \draw [dotted] (0,5) -- (4,5);
      \draw [-latex',densely dotted] (8,1) node [right] {$x_1-x_2=4$}
      .. controls +(180:20pt) and +(-60:20pt) .. (6,2); 
      \foreach \x in {1,2,3,4,5,6,7,8,9}
      {
        \draw (\x,.15) -- (\x,-.15) node [below] {$\x$};
      }
      \draw (10,-.2) node [below] {$x_1$};
      \foreach \y in {1,2,3,4,5}
      {
        \draw (.15,\y) -- (-.15,\y) node [left] {$\y$};
      }
      \draw (-.15,6) node [left] {$x_2$};
      \draw (-.15,-.15) node [left,below] {$0$};
    \end{tikzpicture}}
  \hfill
  \subfigure[DBM $M$, $\val{M} = Z$\label{subfig:M}]{\quad
    \begin{tikzpicture}
      \draw (0,0) node {$\begin{array}{c@{}c}       
                           & \begin{array}{c@{\hspace*{4pt}}c@{\hspace*{4pt}}c} x_0 & x_1 & x_2
                             \end{array} \\
                           \begin{array}{c} x_0 \\ x_1 \\ x_2 \end{array} &
                                                                            \mathrm{\left(\begin{array}{ccc} \infty & -3 &
                                                                                                                           \infty \\ \infty & \infty & 4 \\
                                                                                            5 & \infty & \infty
                                                                                          \end{array}\right)} \\
                           &
                         \end{array}$};
                     \end{tikzpicture}\quad}
                   \hfill
                   \subfigure[Normal form $\phi(M)$\label{subfig:nfM}]{\quad
\begin{tikzpicture}
      \draw (0,0) node {$\begin{array}{c@{\hspace*{4pt}}c}       
    & \begin{array}{c@{\hspace*{4pt}}c@{\hspace*{4pt}}c} x_0 & x_1 & x_2
    \end{array} \\
    \begin{array}{c} x_0 \\ x_1 \\ x_2 \end{array} &
    \mathrm{\left(\begin{array}{c@{\hspace*{7pt}}c@{\hspace*{7pt}}c} 0 & -3 & 0
          \\ 9 & 0 & 4 \\
          5 & 2 & 0
        \end{array}\right)} \\
    &
  \end{array}$};
\end{tikzpicture}\quad}
  \caption{Representations of zone $Z$ defined by
    $\left((x_1 \geq 3)\ \wedge\ (x_2 \leq 5)\ \wedge\ (x_1 - x_2 \leq
      4)\right)$.\label{fig:zone}}
\end{figure}
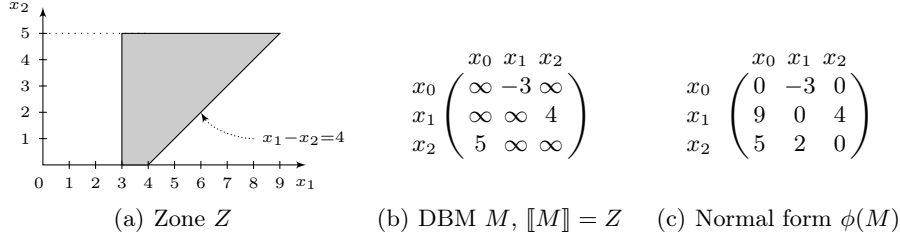

A \emph{zone} over $X$ is a set of clock valuations defined by a
general clock constraint of $\calC(X)$. Figure~\ref{subfig:Z} gives an
example of a zone $Z$ over $X = \{x_1,x_2\}$.

\noindent Given a set of valuations $W$ (over $X$), we define the
following operations:
\begin{itemize}
\item \textit{Intersection of $W$ and $W'$:}
  $W \cap W' = \{v \in \IR_+^X \mid v \in W\ \text{and}\ v \in W'\}$;
\item \textit{Reset to zero of $W$ w.r.t. $Y \subseteq X$:}
  $[Y]W =\{[Y]v \in \IR_+^X \mid v \in W\}$;
\item \textit{Future of $W$:}
  $\overrightarrow{W} = \{v+\delta \in \IR_+^X \mid v\in W\
  \text{and}\ \delta\in \IR_+\}$.
\end{itemize}
One can see that those operations transform a zone into another
zone. Furthermore, if $t = \left(q,g,Y,q'\right) \in T$ is a
transition of $\calA$, then the operator $\Post_t$ defined by
$\Post_t(W) :=\{v' \in \IR_+^X \mid \exists v \in W,\ \exists \delta
\in \IR_+\ \text{s.t.}\ v \models g\ \text{and}\ v'=[Y]v+\delta\}$ can
be computed symbolically as
$\Post_t(W) = \overrightarrow{[Y](W \cap \val{g})}$.  In particular,
if $Z$ is a zone, then so is $\Post_t(Z)$.

A \emph{symbolic state} is a pair $(q,Z)$ where $q\in Q$ is a state
and $Z$ is a nonempty zone.  The \emph{initial} symbolic state is
$s_{0}=(q_0,Z_{0})$ where $Z_{0}=\overrightarrow{\{\mathbf{0}_{X}\}}$.
The set $\nodes$ of \emph{reachable} symbolic states of $\A$ is the
least fixed point of the following rules:
\begin{prooftree}
  \AxiomC{}
  \RightLabel{\scriptsize{Init}}
  \UnaryInfC{$s_{0} \in \nodes$}
\end{prooftree}
\begin{prooftree}
  \AxiomC{$(q,Z)\in\nodes$} \AxiomC{$t = \left(q,g,Y,q'\right) \in T$}
  \AxiomC{$Z' = \Post_t(Z) \neq \emptyset$}
  \RightLabel{\scriptsize{\textsf{Trans}}}
  \TrinaryInfC{$(q',Z')\in\nodes$}
\end{prooftree}

What is called the \emph{forward analysis} is a saturation computation
of the set of reachable states using the above deduction rules.  For
the analysis of this basic forward analysis, details of the
implementation are not very important, but they might be crucial in
practice (see for instance~\cite{BHV00}).  Note that we assume
future-closed zones, since it allows to prove optimality of some of
the abstractions, see~\cite[App.~A]{sri12}.
First, this symbolic forward analysis is sound and complete for
control-state reachability:
\begin{description}
\item[Soundness] If $(q,Z)\in\nodes$, then there exists $v\in Z$
  s.t. $(q,v)$ is reachable in $\A$, i.e., there is a run of $\A$ from
  its initial configuration $(q_{0},\mathbf{0})$ to $(q,v)$.

\item[Completeness] If a configuration $(q,v)$ is reachable in $\A$
  then there is some $(q,Z)\in\nodes$ with $v\in Z$.
\end{description}
The problem with this approach is that the set $\nodes$ may be
infinite, in which case any implementation of this forward analysis
will not terminate on some instances. There are two main approaches to
overcome this problem: \emph{extrapolation} and \emph{simulation}.


\paragraph{Extrapolation.}
The extrapolation approach defines an idempotent operator $\extra$ on
symbolic states: if $(q,Z)$ is a symbolic state then
$\extra\big((q,Z)\big)$ is a symbolic state $(q,Z')$ with
$Z\subseteq Z'$.  This approach changes the \textsf{Trans} rule as
follows:
\begin{prooftree}
  \alwaysNoLine
  \AxiomC{$(q,Z)\in \nodesextra$}
  \AxiomC{$t = \left(q,g,Y,q'\right)\in T$}
  \AxiomC{$(q',Z') = \extra\left((q',\Post_t(Z))\right) \neq \emptyset$}
  \TrinaryInfC{\qquad and there are no $(q',Z'')\in \nodesextra$ with $Z'\subseteq Z''$ \qquad}
  \alwaysSingleLine
  \RightLabel{\scriptsize{$\extra$-\textsf{Trans}}}
  \UnaryInfC{add $(q',Z')$ to $\nodesextra$}
\end{prooftree}

We say that an operator $\extra$ is finite if for every $q \in Q$, for
all infinite sequences $Z_{1},Z_{2},\ldots$ of zones, we find
$1 \le i<j$ with
$\extra\big((q,Z_{j})\big)\subseteq\extra\big((q,Z_{i})\big)$.  An
easy way to ensure finiteness of operator $\extra$ is to ensure that
its range is finite.  If $\extra$ is finite then any forward analysis
induced by the rule $\extra$-\textsf{Trans} always terminates. Note
that the set $\nodesextra$ which is computed may depend on the order
in which the $\extra$-\textsf{Trans} rule is applied.  Optimized
search strategies can be used to efficiently compute
$\nodesextra$~\cite{HT15}.  Since $\Post_t$ is monotone, it is easy to
see that for all $(q,Z)\in\nodes$, there is some
$(q,Z')\in \nodesextra$ with $Z\subseteq Z'$. We deduce that
completeness is preserved by extrapolation.  We discuss in
Section~\ref{sec:extra} extrapolation operators which are finite and
also preserve soundness.  The corresponding forward analysis will
therefore be an algorithm for control-state reachability.


\paragraph{Simulation.}

The simulation approach considers a preorder relation $\preceq$
between symbolic states of $\A$ and restricts the application of the
\textsf{Trans} rule as follows:
\begin{prooftree}
  \alwaysNoLine
  \AxiomC{$(q,Z)\in \nodesprec$}
  \AxiomC{$t = \left(q,g,Y,q'\right)\in T$}
  \AxiomC{$Z' = \Post_t(Z) \neq \emptyset$}
  \TrinaryInfC{\quad and there are no $(q',Z'')\in \nodesprec$ with $(q',Z')\preceq(q',Z'')$ \quad}
  \alwaysSingleLine
  \RightLabel{\scriptsize{$\preceq$-\textsf{Trans}}}
  \UnaryInfC{add $(q',Z')$ to $\nodesprec$}
\end{prooftree}
where $\nodesprec$ is the new set of symbolic states.

We say that $\preceq$ is \emph{finite} if for all infinite sequences
$(q,Z_{1}),(q,Z_{2}),\ldots$ of symbolic states of $\A$, we find
$1 \le i<j$ with $(q,Z_{j})\preceq(q,Z_{i})$.  If $\preceq$ is finite
then the induced forward analysis always terminates.  Note that the
set $\nodesprec$ which is computed may depend on the order in which we
apply the $\preceq$-\textsf{Trans} rule (the optimized search strategy
in~\cite{HT15} can also be applied in this settings).  In all cases,
we have $\nodesprec\subseteq\nodes$, so soundness is preserved.  Now,
if $\preceq$ is a \emph{simulation} (defined in
Section~\ref{sec:simulation}) then completeness is also preserved and
the corresponding forward analysis will therefore be an algorithm for
control-state reachability.

\begin{remark}
  A stronger version of the soundness property is satisfied by the
  exact forward computation based on the \textsf{Trans} rule: if
  $(q,Z)\in\nodes$, then \emph{for all} $v\in Z$, the configuration
  $(q,v)$ is reachable in $\A$. This also holds for the variant based
  on simulation ($\preceq$-\textsf{Trans} rule), but not for the
  variant based on extrapolation (\textsf{extra-Trans} rule).
\end{remark}

The efficiency of the forward analysis crucially depends on the
complexity of applying the \textsf{Trans} rule.  We will see below how
this is implemented in practice.

\subsection{Difference bounded matrices (DBMs)}

The most common data structure for representing zones is the so-called
DBM data structure.  This data structure has been first introduced
in~\cite{BM83} and then set in the framework of timed automata
in~\cite{dill89}.  Several presentations of this data structure can be
found in the literature, for example in~\cite{CGP99,BY04,bouyer04}.

A \emph{difference bounded matrix} (DBM in short) for a set
$X = \{x_1,\dots,x_n\}$ of $n$ clocks is an $(n+1)$-square matrix of
pairs
\[
  (\triangleleft,m) \in \mathbb{V} := (\{<,\leq\} \times \IZ) \cup
  \{(<,\infty)\}.
\]
A DBM
$M = \big((\triangleleft_{i,j},m_{i,j})\big)_{0 \leq i,j \leq n}$
defines the zone:
\[
  \val{M} := \left\{v\colon X \rightarrow \IR_+ \mid \forall\ 0 \leq i,j
    \leq n,\ \overline{v}(x_i) - \overline{v}(x_j) \triangleleft_{i,j} m_{i,j}
  \right\}
\]
where $\overline{v} \in \IR_+^{\{x_0\} \cup X}$ is such that
$\overline{v}_{|X} = v$ and $\overline{v}(x_0)=0$, and where
$\gamma < \infty$ simply means that $\gamma \in \IR_+$.  To simplify
the notations, we assume from now on that all constraints are
non-strict (except $(<,\infty)$), so that coefficients of DBMs can
simply be seen as elements of $\IZ \cup \{\infty\}$.
With this convention, the zone $Z$ of Figure~\ref{subfig:Z} can be
represented by the DBM of Figure~\ref{subfig:M}.

A zone can have several representations using DBMs. For example, the
previous zone can equivalently be represented by the DBM given in
Figure~\ref{subfig:nfM}. This DBM contains constraints that were
implicit in the former representation: for instance, the constraint
$x_1-x_0 \le 9$ encoded in the DBM is implied by $x_1-x_2 \le 4$ and
$x_2-x_0 \le 5$.

With every DBM $M$ we associate its adjacency graph $G_M$, and when
there is no negative cycle in $G_M$, we let $\phi(M)$ be the DBM
obtained by computing the shortest paths in $G_M$ (for instance using
the Floyd-Warshall algorithm). Then $\phi(M)$ is the
smallest\footnote{For the partial order $\le$ defined pointwise.}  DBM
representing the same zone as $M$. It is called the \emph{normal form}
of $M$ and the computation of $\phi(M)$ from $M$ is called
\emph{normalization}. We can then notice the following:
\begin{itemize}
\item $\val{M} \subseteq \val{M'}$ iff $\phi(M) \le M'$ iff $\phi(M)
  \le \phi(M')$;
\item $\val{M} = \emptyset$ iff there is a negative cycle in $G_M$.
\end{itemize}
The first point says that inclusion of zones can be checked
efficiently, i.e., in time $\mathcal{O}(|X|^{2})$, on normal forms of
DBMs.  The second point says that emptiness can be checked in time
$\mathcal{O}(|X|^{3})$; this check will never be used as is, and
emptiness is detected while computing the various operations (keeping
DBMs in normal form).

Finally, the various operations on zones that we need for forward
analysis can be done efficiently on DBMs:
\begin{itemize}
  
\item \textit{Intersection of $M$ and $M'$:} $\min(M,M')$.  The
  complexity is $\mathcal{O}(|X|^{2})$.  But notice that, even if we
  start with DBMs $M,M'$ in normal form, the resulting DBM
  $\min(M,M')$ is not necessarily in normal form. To get the result in
  normal form, we need normalization which takes time
  $\mathcal{O}(|X|^{3})$.
  
\item \textit{Intersection of $M$ with an atomic constraint $g$ of the
    form $x_{i}-x_{j}\leq c$:} assuming $M=\phi(M)$ is in normal form,
  the normal form $M'$ of $M\cap g$ is defined by
  $m'_{k,\ell}=\min(m_{k,\ell},m_{k,i}+c+m_{j,\ell})$ and can be
  computed in time $\mathcal{O}(|X|^{2})$. When $g$ is an arbitrary
  constraint, we repeat the above operation with every atomic
  constraints in $g$.  The complexity is
  $\mathcal{O}(|X|^{2}\cdot|g|)$.

\item \textit{Reset $x_i$ to zero in $M$:} as $x_i=x_0$ after the
  reset, this is achieved on $\phi(M)$ as follows: for all $j$, set
  $m_{i,j}$ to $m_{0,j}$, and set $m_{j,i}$ to $m_{j,0}$.  The
  resulting DBM is still in normal form.  The complexity is
  $\mathcal{O}(|X|)$.
 
  Resetting a set $Y$ of clocks in $M$ amounts to repeat the above
  operation for all $x_i\in Y$.  The complexity is
  $\mathcal{O}(|X|\cdot|Y|)$.
\item \textit{Future of $M$:} on $\phi(M)$, relax (i.e., set to
  $\infty$) all upper-bound constraints $x_{i}-x_{0}$ with $i\neq0$
  (that is, all coefficients on column $x_0$ except the first one).
  The resulting DBM is still in normal form.  The complexity is
  $\mathcal{O}(|X|)$.
\end{itemize}

\subsection{Efficiency of the forward analysis}
\label{subsec:efficiency}

Let us discuss the complexity of applying the \textsf{Trans} rule.
First, given a transition $t=(q,g,Y,q')$ and a zone $Z$ represented by
a DBM $M$ in normal form, we compute a DBM $M'$ in normal form for the
zone $\Post_{t}(Z)$. To do so, we intersect $M$ with $g$, then we
reset clocks in $Y$ and finally we let time elapse (future). As
explained above, this is computed in time
$\mathcal{O}(|X|^{2}\cdot|t|)$.

We will define in Section~\ref{sec:extra} extrapolation operators
$\extra(q,Z)=(q,Z')$ which are finite and sound for diagonal-free
timed automata.  When $Z$ is represented by a DBM $M$ in normal form,
we compute a DBM $M'$ for $Z'$ in time $\mathcal{O}(|X|^{2})$.  But
$M'$ is not in normal form, so we need normalization, which takes time
$\mathcal{O}(|X|^{3})$.  We saw above that the inclusion test
$Z'\subseteq Z''$ required by the \textsf{extra-Trans} rule can be
performed in time $\mathcal{O}(|X|^{2})$ when the DBM for $Z'$ is in
normal form.  Hence, we may also apply the \textsf{extra-Trans} rule
efficiently.

In Section~\ref{sec:simulation}, we will define a preorder $\preceq$
which is a finite simulation relation, for all timed automata, even
those using diagonal constraints.  Moreover, the test
$(q,Z)\preceq(q,Z')$ can be checked in time $\mathcal{O}(|X|^{2})$
when the timed automaton is diagonal-free.  Hence, we may apply the
$\preceq$-\textsf{Trans} rule of the forward analysis efficiently.  If
diagonal constraints are allowed, checking the simulation
$(q,Z)\preceq(q,Z')$ is an NP-complete problem.

Notice that, when applying the \textsf{extra-Trans} rule (resp. the
$\preceq$-\textsf{Trans} rule), we need to check inclusion
$Z'\subseteq Z''$ (resp. simulation $(q',Z')\preceq(q',Z'')$) against
all already computed zones $Z''$ with the same state.  Hence, checking
simulation (resp.  inclusion) is the dominant operation when applying
the \textsf{Trans} rule.

\section{Extrapolation: a first solution}\label{sec:extra}

\subsection{The first extrapolation operator}

We discuss here the first extrapolation operator which has been
defined~\cite{DT98} to ensure good properties of the forward analysis,
and which has been prevalent in the timed systems verification
community until the early 2010's.

Let $K \in \IN^X$ be a tuple of integers.  A $K$-bounded clock
constraint is a clock constraint where clock $x$ is only compared to
constants between $-K_x$ and $+K_x$, and the difference $x - y$ is
compared to constants between $-K_y$ and $+K_x$. By extension, a
$K$-bounded zone is a zone which can be defined by a $K$-bounded clock
constraint. If $(q,Z)$ is a symbolic state, the
\emph{$K$-extrapolation} $\extra_K\big((q,Z)\big)$ of $(q,Z)$ is the
pair $(q,Z')$ such that $Z'$ is the smallest $K$-bounded zone which
contains $Z$.  Intuitively, this operator relaxes constraints bounding
clock $x$ where constants are larger than $K$.

This operation is well-defined on DBMs: if $M$ is a DBM in normal form
representing $Z$, a DBM representing the zone $Z'$ such that
$(q,Z') = \extra_K\big((q,Z)\big)$ is $M'$ (denoted $\extra_K(M)$)
such that:
\[
(\triangleleft'_{i,j};m'_{i,j}) := \left\{\begin{array}{ll}
  (<;\infty) & \text{if}\ m_{i,j}>K_{x_i} \\
  (<;-K_{x_j}) & \text{if}\ m_{i,j}<-K_{x_j} \\
(\triangleleft_{i,j,},m_{i,j}) & \text{otherwise}
\end{array}\right.
\]

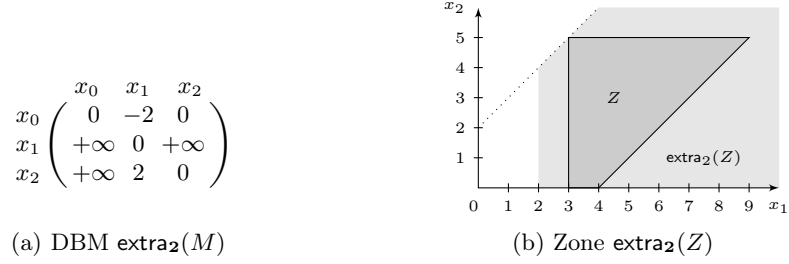
\begin{figure}[t]
  \subfigure[DBM $\extra_{\mathbf{2}}(M)$\label{subfig:extraM}]{
    \begin{tikzpicture}
\path [use as bounding box] (-3,-1) -- (3,1.5);
      \draw (0,0) node {$\begin{array}{c@{}c}       
                           & \begin{array}{c@{\hspace*{4pt}}c@{\hspace*{4pt}}c} x_0 & ~~x_1 & ~~x_2
                             \end{array} \\
                           \begin{array}{c} x_0 \\ x_1 \\ x_2 \end{array} &
                                                                            \mathrm{\left(\begin{array}{ccc} 0 & -2 & 0 \\
                                                                                            +\infty
                                                                                                               &
                                                                                                                 0 & +\infty \\
                                                                                            +\infty & 2 & 0
                                                                                          \end{array}\right)} \\
                           &
                         \end{array}$};
                     \end{tikzpicture}} \hfill \subfigure[Zone
                   $\extra_{\mathbf{2}}(Z)$\label{subfig:extraZ}]{
    \begin{tikzpicture}[scale=.4]
      \everymath{\scriptstyle}
      \fill [fill=black!10!white] (2,0) -- (10,0) -- (10,6) -- (4,6) -- (2,4) -- cycle;
      \draw [fill=black!20!white] (3,0) -- (4,0) -- (9,5) -- (3,5) -- cycle;
      \draw [latex'-latex'] (0,6) -- (0,0) -- (10,0);
      \draw [dotted] (0,2) -- (4,6); 
      \foreach \x in {1,2,3,4,5,6,7,8,9}
      {
        \draw (\x,.15) -- (\x,-.15) node [below] {$\x$};
      }
      \draw (10,-.2) node [below] {$x_1$};
      \foreach \y in {1,2,3,4,5}
      {
        \draw (.15,\y) -- (-.15,\y) node [left] {$\y$};
      }
      \draw (-.15,6) node [left] {$x_2$};
      \draw (-.15,-.15) node [left,below] {$0$};
      \draw (4.5,3) node {$Z$};
      \draw (7.5,1) node {$\extra_2(Z)$};
    \end{tikzpicture}}
  \hfill
  \caption{Illustration of the extrapolation operator}
\end{figure}

Considering the zone given in Figure~\ref{fig:zone}, its extrapolation
w.r.t. $\extra_{\mathbf{2}}$ (where $\mathbf{2}$ denotes the tuples
associating $2$ to every clock) is the DBM of
Figure~\ref{subfig:extraM} (which is not in normal form) and is
depicted in Figure~\ref{subfig:extraZ}.

Obviously, $\extra_K$ is finite (since its range is finite), hence any
forward analysis using the $K$-extrapolation will terminate.
Thus only soundness requires some careful proof, and will actually not
hold in general. The first complete proof of soundness was given
in~\cite{bouyer03,bouyer04}, and was done in the more general context
of updatable timed automata. It requires to show that the
extrapolation is a \emph{simulation-based abstraction}, that is, there
is some simulation relation (in a sense that we will make clear in the
next section) such that any configuration which is computed in
$\nodesextra$ can be simulated by some configuration in $\nodes$.

\begin{theorem}
  Let $K \in \IN^X$ be a tuple of integers and let $\calA$ be a
  $K$-bounded\footnote{for every $x \in X$, for every constraint
    $x \bowtie c$ appearing in $\calA$, $K_x \ge c$.}  and
  diagonal-free timed automaton.  Then, the forward analysis which
  computes $\calS_{\extra_K}$ terminates, and is sound and complete.
\end{theorem}

\subsection{Two refinements of this approach}

\subsubsection{State-dependent constants.}
\label{subsec:state}
The extrapolation operator is parametrized using a tuple
$K \in \IN^X$. This can actually be refined and made state-dependent,
by considering one constant per state and per clock (i.e., a tuple
$K \in \IN_\infty^{X \times Q}$, with
$\IN_\infty = \IN \cup \{+\infty\}$), taking only into account
constraints which have an impact on the current state: for instance,
if $x$ is compared to constant $c$ and then reset before reaching
state $q$, then constant $c$ is not taken into account (this limited
propagation idea will be presented in the context of $\Gg$-simulation
in Section~\ref{sec:simulation}). This refined extrapolation was
proven sound for forward analysis in~\cite{BBFL03}.

Note that the notion of active and inactive clocks of~\cite{DT98} can
be recovered from this refined extrapolation: a clock $x$ is inactive
at state $q$ whenever the constant $K_{x,q}$ is $+\infty$, in which
case it can be ignored at state $q$.

\subsubsection{Distinguishing lower and upper bounds.}
\label{sec:extraLU}
Another refinement, developed in~\cite{BBLP04}, consists in
distinguishing lower-bounding and upper-bounding constraints of
clocks: instead of having one constant per clock, we consider two
constants $L,U \in \IN_\infty^X$. An $LU$-bounded zone is one where
$x$ is compared to constants between $-U_x$ and $+L_x$, and the
difference $x - y$ is compared to constants between $-U_y$ and
$+L_x$. Then the \emph{$LU$-extrapolation}
$\extra_{LU}\big((q,Z)\big)$ of $(q,Z)$ is the pair $(q,Z')$ such that
$Z'$ is the smallest $LU$-bounded zone that contains $Z$.

This operation can be better understood on DBMs: if $M$ is a DBM in
normal form representing $Z$, a DBM representing the zone $Z'$ such
that $(q,Z') = \extra_{LU}\big((q,Z)\big)$ is $M'$ (denoted
$\extra_{LU}(M)$) such that:
\[
(\triangleleft'_{i,j};m'_{i,j}) := \left\{\begin{array}{ll}
  (<;\infty) & \text{if}\ m_{i,j}>L_{x_i} \\
  (<;-U_{x_j}) & \text{if}\ m_{i,j}<-U_{x_j} \\
(\triangleleft_{i,j,},m_{i,j}) & \text{otherwise}
\end{array}\right.
\]
Obviously, $\extra_{LU}$ is finite (since its range is finite), hence
any forward analysis using the $LU$-extrapolation will terminate.
Only soundness requires some careful proof, see~\cite{BBLP04}.  It
uses a $LU$-simulation, which is a technical notion that we do not
discuss here, since a refinement will be discussed in
Section~\ref{sec:simulation}.

\begin{theorem}
  Let $L,U \in \IN^X$ be tuples of integers and let $\calA$ be a
  diagonal-free and $LU$-bounded\footnote{for every $x \in X$, for
    every constraint $x < c$ or $x \le c$ (resp.\ $x>d$ or $x \ge d$)
    appearing in $\calA$, $U_x \ge c$ (resp.\ $L_x \ge d$).} timed
  automaton.  Then the forward analysis which computes
  $\calS_{\extra_{LU}}$ terminates, and is sound and complete.
\end{theorem}
Further refinements can be made, see~\cite{BBLP04}, but will not be
discussed here.

\subsection{Problems with diagonal clock constraints}
\label{sec:probl-with-extra}

Surprisingly, the approach using extrapolation is not appropriate to
analyze general timed automata, which may use diagonal clock
constraints, and until recent advances presented in
Section~\ref{sec:simulation}, there was no really satisfactory
solution for analyzing general timed automata, see~\cite{BLR05}.

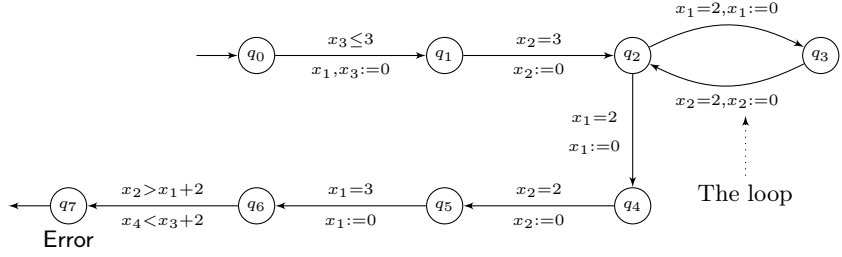
\begin{figure}[t]
  \begin{center}
    \begin{tikzpicture}
      \everymath{\scriptstyle}
      \draw (0,0) node [draw,circle,inner sep=2pt] (A) {$q_0$};
      \draw (2.5,0) node [draw,circle,inner sep=2pt] (B) {$q_1$};
      \draw (5,0) node [draw,circle,inner sep=2pt] (C) {$q_2$};
      \draw (7.5,0) node [draw,circle,inner sep=2pt] (D) {$q_3$};
      \draw (5,-2) node [draw,circle,inner sep=2pt] (E) {$q_4$};
      \draw (2.5,-2) node [draw,circle,inner sep=2pt] (F) {$q_5$};
      \draw (0,-2) node [draw,circle,inner sep=2pt] (G) {$q_6$};
      \draw (-2.5,-2) node [draw,circle,inner sep=2pt] (H) {$q_7$} node [below=6pt] {{\footnotesize \textsf{Error}}};
      \draw [latex'-] (A) -- (-.8,0);
      \draw [-latex'] (A) -- (B) node [midway,above] {$x_3 \leq 3$} node [midway,below] {$x_1,x_3:=0$};
      \draw [-latex'] (B) -- (C) node [midway,above] {$x_2=3$} node [midway,below] {$x_2:=0$};
      \draw [-latex'] (C) .. controls +(1,.5) and +(-1,.5) .. (D) node [midway,above] {$x_1=2,x_1:=0$};
      \draw [-latex'] (D) .. controls +(-1,-.5) and +(1,-.5) .. (C) node [midway,below] {$x_2=2,x_2:=0$};
      \draw [-latex'] (C) -- (E) node [midway,left] {$\begin{array}{c}
                                                        x_1=2 \\
                                                        x_1:=0 \end{array}$};
      \draw [-latex'] (E) -- (F) node [midway,above] {$x_2=2$} node [midway,below] {$x_2:=0$};
      \draw [-latex'] (F) -- (G) node [midway,above] {$x_1=3$} node [midway,below] {$x_1:=0$};
      \draw [-latex'] (G) -- (H) node [midway,above] {$x_2>x_1+2$} node [midway,below] {$x_4<x_3+2$};
      \draw [-latex'] (H) -- +(-.8,0);

      \draw [-latex',dotted] (6.5,-1.6) node [below] {{\footnotesize The loop}} -- +(0,.8);
    \end{tikzpicture}
  \end{center}
  \caption{Automaton $\A_{\textsf{bug}}$, with diagonal constraints,
    which cannot be analyzed by any algorithm using a zone
    extrapolation\label{fig:bug}}
\end{figure}

Indeed, consider the timed automaton $\A_{\textsf{bug}}$ depicted on
Figure~\ref{fig:bug}. Then it is proven in~\cite{bouyer04} that it
cannot be analyzed using a forward analysis computation using any
finite-ranged extrapolation (like those we have described before).

\begin{proposition}
  There is no extrapolation operator $\extra$ with finite range, for
  which the forward analysis which computes $\calS_{\extra}$ is sound
  on $\A_{\textsf{bug}}$. This obviously applies to operators
  $\extra_K$ and $\extra_{LU}$ that we have discussed before.
\end{proposition}

Let us explain what happens in this timed automaton. 
If the loop is taken $\alpha$ times, the zone which is reachable at
$q_6$ is $Z_\alpha$, defined by the constraint
\[
  (1 \leq x_2 - x_1 \leq 3) \wedge (1 \leq x_4 - x_3 \leq 3) \wedge (
  x_4 - x_2 = x_3 - x_1 = 2 \alpha + 5).
\]
If we apply an extrapolation operator, $\extra_K$ or $\extra_{LU}$, to
this zone when $\alpha$ is large, we will lose the property that
$x_2-x_1 = x_4-x_3$ in $Z_\alpha$ (this implicit constraint prevents
the transition $q_6 \to q_7$ to be taken). The saturation algorithm
which computes $\nodesextra$ will then add some symbolic state of the
form $(q_7,Z)$ to $\nodesextra$, even though control-state $q_7$ is
not reachable!


\section{Simulation: an alternative solution}\label{sec:simulation}

The extrapolation approach is lucrative as it allows to work with
zones and efficient operations over them. There are two main
shortcomings: the approach does not work when diagonal constraints are
present, and secondly even for diagonal-free automata, it does not use
the coarsest known abstraction of zones, which was proposed
in~\cite{BBLP04}. An $LU$-simulation relation $v \lu v'$ between
valuations was defined in ~\cite{BBLP04}. In principle, one could
consider an extrapolation
$\extra(Z) = \{ v \in \IR_+^X \mid \exists v' \in Z \text{ with } v
\lu v'\}$, which is simply the downward closure of $Z$ with respect to
the $LU$-simulation. However, there are zones for which this downward
closure is not even convex. Hence the $LU$-simulation cannot be used
with the extrapolation approach. The coarser the abstraction, the
smaller is the number of zones enumerated and therefore it is
motivating to look for ways to use coarser abstractions. These two
questions have been the driving force of research in this area over
the last decade.

The central idea is to use simulations between zones, instead of
extrapolating them. Extrapolations need to store abstracted zones
explicitly. This is a bottleneck for the usage of non-convex
abstractions. The simulation approach, in principle, eliminates this
problem. The missing piece is an efficient simulation test between
zones. This was settled in~\cite{HSW12} for diagonal-free automata
where it was shown that $Z \lu Z'$ (the inclusion up to
$LU$-simulation, more precisely
$\forall v \in Z, \exists v' \in Z' \text{ s.t. } v \lu v'$) can be
done in time quadratic in the number of clocks, as efficient as a
plain inclusion $Z \subseteq Z'$.  Later, a new simulation, known as
the $\Gg$-simulation, was developed for automata with diagonal
constraints~\cite{GMS19}.  It is based on a map $\Gg$ associating sets
of constraints to states of the timed automaton.  It was not as pretty
a picture as in the diagonal-free case since the corresponding
simulation test was shown to be NP-complete (\cite{GMS19-arXiv} and
\cite[Section 4.4]{Sayan}). In spite of this, the $\Gg$-simulation
approach is promising: (1) when restricted to diagonal-free automata,
the $\Gg$-simulation is coarser than the $LU$-simulation and the
simulation test remains quadratic~\cite{GMS19-arXiv} (see
\cite[Section 4.2]{Sayan} for a more elaborate exposition), (2) in the
general case when diagonals are present, heuristics have been proposed
to get a fast test in practice and have been shown to work well in
experiments and (3) the approach can be extended seamlessly for
automata with updates.  Previous approaches for dealing with diagonals
had to eliminate the diagonal constraints explicitly and therefore
resulted in a blow-up in space. Here the difficulty gets absorbed into
the time to do the simulation test.  For automata with updates, no
zone based solution was available previously.

In this section, we present the $\Gg$-simulation framework. Recall
from Section~\ref{sec:symbolic} that the simulation approach makes use of a preorder $\preceq$
to prune the saturation computation of the symbolic states. More
specifically, the $\preceq$-\textsf{Trans} rule adds a new symbolic
state $(q', Z')$ if there are no existing symbolic states $(q', Z'')$
with $(q', Z') \preceq (q', Z'')$. The idea is that if
$(q', Z') \preceq (q', Z'')$ then every control state reachable from
$(q', Z')$ should be reachable from $(q', Z'')$ and therefore it is
not necessary to explore further from $(q', Z')$. This idea is neatly
captured by simulation relations. 

For the rest of this section, we fix a timed automaton $\A$.

  \begin{definition}[Simulation]\label{def:simulation}
    A preorder relation $\preceq$ between configurations of $\calA$
    having the same control state is said to be a simulation if for
    every $(q,v) \preceq (q, v')$, two properties are satisfied:
    \begin{enumerate}
      \item[(1)] (time elapse): for every $\delta\in \IR_+$, we have 
      $(q, v+\delta) \preceq (q, v' + \delta)$, and
    \item[(2)] (discrete transition): for every
      $(q, v) \xra{t} (q_1, v_1)$ with $t$ a transition of $\A$, there
      exists $(q, v') \xra{t} (q_1, v'_1)$ with
      $(q_1, v_1) \preceq (q_1, v'_1)$.
    \end{enumerate}
    We extend $\preceq$ to zones by defining $(q, Z) \preceq (q, Z')$
    if for every $v \in Z$ there exists $v' \in Z'$ such that
    $(q, v) \preceq (q, v')$.
  \end{definition}
  
  Notice that if $(q,v)\preceq(q,v')$ and there is a run from $(q,v)$
  to $(q_{1},v_{1})$ alternating time elapses and discrete
  transitions, then there is a run from $(q,v')$ to some
  $(q_{1},v'_{1})$ using the same sequence of time elapses and
  discrete transitions.

  The relation as defined above is known as a strong-timed simulation
  in literature~\cite{TY01} since we require the same $\delta$ from
  $(q, v')$ in condition (1). When we relax (1) to say that we have
  $(q, v+\delta) \preceq (q, v' + \delta')$ for some
  $\delta' \in \IR_+$, the relation is called a time-abstract
  simulation. The concrete relations that we know so far turn out to
  be strong-timed simulations and so we restrict to the above
  definition.  Our goal is to now find a concrete simulation
  relation. 

  From condition (2) of Definition~\ref{def:simulation}, it is clear
  that the guards of the automaton play a crucial role in coming up
  with a simulation relation. So, as a first step, we collect
  a set of \emph{relevant} constraints at each state of the
  automaton. Clearly, the guards from the outgoing transitions
  of a state $q$ are relevant at $q$. Now, consider a transition $q
  \xra{t} q_1$ of the automaton. A constraint that is relevant at
  $q_1$ needs to be propagated in some form to $q$ depending on the
  resets at $t$.  Otherwise,
 there is no means to achieve 
  the forward propagation of the simulation (condition (2)). The next
  definition formalizes this idea through a fixpoint characterization.
  
  \begin{definition}[Constraint map, \cite{GMS19,Sayan}]\label{def:g-map}
    The constraint map $\Gg$ of $\A$ associates a set $\Gg(q)$ of
    constraints to every state $q\in Q$.  It is obtained as the least
    fixpoint of the following system of equations:
    \begin{align*}
      \Gg(q) = \bigcup_{(q, g, Y, q') \in T} 
      \left(
      \{\text{ atomic constraints of $g$ }\} ~\cup~ 
      \bigcup_{\varphi \in \Gg(q')} \pre(\varphi, Y)
      \right)
    \end{align*}
    where $\pre(x \bowtie c, Y)$ is $\{x \bowtie c\}$ if $x \notin Y$
    and is the empty set when $x \in Y$, and $\pre(x - y \bowtie c)$
    is $\{x \bowtie c\}$ if $x \notin Y, y \in Y$; is
    $\{-y \bowtie c\}$ if $x \in Y, y \notin Y$; is
    $\{x - y \bowtie c\}$ when $x, y \notin Y$, and is the empty set
    when both $x, y \in Y$.
  \end{definition}

  For the automaton $\A_{\textsf{bug}}$ in Figure~\ref{fig:bug}, we
  have $\Gg(q_6) =\{ x_2 - x_1 > 2, x_4 - x_3 < 2\}$,
  $\Gg(q_2) = \{x_1 = 2, x_2 = 2, x_4 - x_3 < 2 \}$ and
  $\Gg(q_0) = \{x_3 \le 3, x_2 = 3, x_4 < 2 \}$. To see why $\Gg(q_0)$
  is this set of constraints, notice that $x_3 \le 3$ appears in the
  guard out of $q_0$. Since $x_1$ and $x_3$ are reset in
  $q_0 \xra{} q_1$, constraints on these clocks which appear later are
  not relevant at $q_0$ (and will be made empty set by the $\pre$
  operator). Since $x_2$ is reset in $q_1 \xra{} q_2$, only the
  $x_2 = 3$ propagates to $q_0$. Finally, the constraint
  $x_4 - x_3 < 2$ in $q_6 \xra{} q_7$ propagates all the way back to
  $q_1$ as it is, and due to the reset of $x_3$ in $q_0 \xra{} q_1$,
  we get $\pre(x_4 - x_3 < 2, \{x_1, x_3\}) = \{x_4 < 2\}$ and hence
  we have $x_4 < 2$ in $\Gg(q_0)$.
 
  The computation of the constraint map of a timed automaton is in the
  same spirit as the local computation of constants proposed for the
  extrapolation operator in~\cite{BBFL03} (mentioned in
  Section~\ref{subsec:state}), although now we deal with constraint
  sets instead of constants.  We now have the parameters for the
  simulation ready.  Our next definition gives a relation between
  configurations that is based on these computed constraints.
  
  \begin{definition}[$\Gg$-preorder, \cite{GMS19,Sayan}]
    Let $\Gg$ be the associated constraint map to $\A$.  We define
    $(q, v) \preceq_{\Gg} (q, v')$ if for all $\delta \in \IR_+$ and
    all $\varphi \in \Gg(q)$, $v + \delta \models \varphi$ implies
    $v' + \delta \models \varphi$.
  \end{definition}

  Thanks to its definition, it is not difficult to show that the
  $\Gg$-preorder $\preceq_{\Gg}$ is a simulation relation.  Showing
  finiteness is more involved.
  Let $Z$ be a zone and $q$ be a state of $\A$. We write
  ${\downarrow}_{q,\Gg}Z=\{v \in \IR_+^X \mid \exists v'\in Z \text{
    s.t.\ } (q,v) \preceq_{\Gg} (q, v')\}$.  Let $M$ be the maximum
  constant appearing in $\Gg(q)$.  From $M$, one can construct a
  finite partition of the space of valuations and show that
  ${\downarrow}_{q,\Gg}$ is a union of its classes. Therefore, the set
  $\{ {\downarrow}_{q,\Gg} Z \mid Z \text{ is a zone } \}$ is finite,
  which implies that the simulation $\preceq_{\Gg}$ is
  finite~\cite{GMS19-arXiv}, \cite[Section 4.1]{Sayan}.
  
  \begin{theorem}[\cite{GMS19-arXiv,Sayan}]
    Let $\A$ be a timed automaton possibly with diagonal
    constraints. The relation $\preceq_{\Gg}$ is a finite simulation
    relation.  Moreover the forward analysis which computes
    $\calS_{\preceq_{\Gg}}$ terminates, and is sound and complete.
  \end{theorem}

  Coming back to the example $\A_{\textsf{bug}}$, let us see how the
  simulation approach works here. Since no new valuations are added
  (in contrast to the extrapolation approach), soundness is always
  guaranteed and $q_7$ would be unreachable. The question is about
  termination. Due to the loop, there are potentially infinitely many
  zones appearing at state $q_{6}$.  As seen in
  Section~\ref{sec:probl-with-extra}, the zone $Z_{\alpha}$ reached at
  $q_6$ after $\alpha$ iterations of the loop satisfies
  $x_4 - x_2 = x_3 - x_1 = 2\alpha + 5$ and
  $1 \leq x_4 - x_3 = x_2 - x_1 \leq 3$.  We claim that
  $(q_{6},Z_{\alpha}) \preceq_{\Gg} (q_{6},Z_{\beta})$ for all
  $\alpha,\beta\geq0$. Indeed, from $v\in Z_{\alpha}$ we define $v'$
  by $v'(x_{1})=v(x_{1})$, $v'(x_{2})=v(x_{2})$,
  $v'(x_{3})=v(x_{3})+2(\beta-\alpha)$ and
  $v'(x_{4})=v(x_{4})+2(\beta-\alpha)$. Intuitively, $v'$ is obtained
  by taking the loop $\beta$ times instead of $\alpha$ times and
  keeping the other delays unchanged.  We can check that
  $v'\in Z_{\beta}$. Moreover, $v'(x_{2})-v'(x_{1})=v(x_{2})-v(x_{1})$
  and $v'(x_{4})-v'(x_{3})=v(x_{4})-v(x_{3})$.  As we have seen above,
  $\Gg(q_6) = \{x_2 - x_1 > 2, x_4 - x_3 < 2\}$.  We deduce that
  $(q_{6},v)\preceq(q_{6},v')$.  Hence the zone enumeration will stop
  at the second zone appearing at $q_6$, due to the simulation. An
  analogous situation happens at other states $q_2, q_3, q_4$ and
  $q_5$.

  When $\Gg(q)$ contains no diagonal constraints, the test
  $(q, Z) \preceq_{\Gg} (q, Z')$ can be done in time quadratic in the
  number of clocks~\cite{GMS19-arXiv}, \cite[Sec.\ 4.2]{Sayan}. The
  main idea is that this test can be broken into tests of the form
  $Z \preceq_{\{x \lhd c, y \rhd d\}} Z'$ where $x \lhd c$ with
  $\lhd \in \{<, \le\}$ is an upper bound constraint and $y \rhd d$
  with $\rhd \in \{ >, \ge \}$ is a lower bound constraint in
  $\Gg(q)$.  When $\Gg(q)$ contains diagonal constraints, the test is
  NP-complete~\cite{GMS19-arXiv},\cite[Section 4.4]{Sayan}. More
  precisely, the test is exponential in the number of diagonals
  present in the set $\Gg(q)$. An algorithm to compute this test in
  the general case appears in \cite{GMS19}. The algorithm essentially
  reduces $(q, Z) \preceq_{\Gg} (q, Z')$ to several (exponentially
  many in the worst case) simulation checks over non-diagonal
  constraints.  It employs heuristics which can reduce this number in
  practice.
  
  \subsubsection{Dealing with updates.}

  The reset operation can be extended with updates to clocks. An
  update $up$ to the set of clocks $X$ is a function which maps each
  clock $x$ to an expression $x := c$ or $x: = y + d $ where
  $c \in \mathbb{N}$, $y \in X$ (which could as well be $x$) and
  $d \in \mathbb{Z}$. Automata with such update functions are known as
  updatable timed automata~\cite{BDFP04}. Due to the presence of
  updates $x:= x +1$ and $x: = x - 1$, reachability becomes
  undecidable, even with $0$ time elapse. Several decidable subclasses
  have been studied by constructing a finite region equivalence.

  The zone based simulation method can be extended to updatable timed
  automata~\cite{GMS19}, just by changing one definition: in
  Definition~\ref{def:g-map}, replace transition $(q, g, R, q')$ with
  $(q, g, up, q')$ and the the operator $\pre(\varphi, R)$ with an
  operation $\pre(\varphi, up)$ that is defined as follows. Define
  $up_x$ to be $c$ (resp.\ $y + d$) if $up$ maps $x$ to $x := c$
  (resp.\ $x := y +d $). The constraint $\pre(\varphi, up)$ is
  obtained by replacing occurrence of $x$ in $\varphi$ with
  $up_x$. For example $\pre(x - y \le 5, x:= z - 2)$ is
  $\{z - y \le 7\}$. The least fixpoint of the equations in
  Definition~\ref{def:g-map} may not be finite. When it is finite, the
  constraint map $\Gg$ and the $\Gg$-preorder can be used to get a
  finite simulation as before. An algorithm to detect termination of
  the fixpoint computation is provided.  For all decidable subclasses
  tabulated in \cite{BDFP04}, the constraint map $\Gg$ is finite. A
  refined version of the constraint map computation that uses an
  operator $\pre(\varphi, g, up)$ taking the guard $g$ into
  consideration gives finite constraint maps for a larger class of
  automata, including timed automata with bounded subtraction that has
  been used to model preemptive scheduling~\cite{FKP+07}.

\section{Beyond reachability}
\label{sec:more}

\subsection{Weighted timed automata}

Timed automata do not offer the possibility to model other quantities
than time (or durations). It may however be useful to also model other
quantities. The more general model of hybrid automata is unfortunately
not adequate for automatic verification, since it is undecidable in
general~\cite{HKPV95}. In 2001, the model of weighted timed automata
(also called priced timed automata at that time) has been
proposed~\cite{ATP01,BFH+01}: on top of a standard timed automaton, a
weight is associated with every state (where it is a rate) and with
every transition (where it is a discrete change). Let
$\wgt \colon Q \cup T \to \mathbb{Z}$ be such a weight function. It
allows to give a \emph{cost} value to time elapse
$(q, v) \xra{\delta} (q, v + \delta)$ (with $\delta \in \IR_+$) as
$\delta \cdot \wgt(q)$, and to discrete transition
$(q,v) \xra{t} (q_1, v_1)$ as $\wgt(t)$. The cost is then defined on a
finite run by summing up the costs of all single steps (time elapses
or discrete transitions) along it.  This cost function, defined on
every finite run, represents the evolution of an observer variable
which is piecewise-linear w.r.t. time elapsing, and it can be used for
various purposes: ensure that it remains within given
bounds~\cite{BFLMS08,BFLM10,BLM12}; or optimize its value while
ensuring to reach some state~\cite{ATP01,BFH+01,BBBR07}. The model is
discussed in the survey~\cite{BFLM11}. Here, we focus on this last
problem, assuming that the weight function is nonnegative.

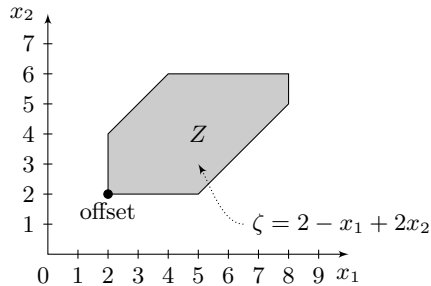
\begin{figure}[t]
  \begin{center}
    \begin{tikzpicture}[scale=.4]
      \draw [latex'-latex'] (0,8) -- (0,0) -- (10,0);    
      
      \draw [fill=black!20!white] (2,2) --(5,2) -- (8,5) -- (8,6) --
      (4,6) --(2,4) --cycle;
      \draw [-latex',densely dotted] (6.5,1) node [right] {$\zeta = 2-x_1+2x_2$}
      .. controls +(180:20pt) and +(-60:20pt) .. (5,3); 
      \foreach \x in {1,2,3,4,5,6,7,8,9}
      {
        \draw (\x,.15) -- (\x,-.15) node [below] {$\x$};
      }
      \draw (10,-.2) node [below] {$x_1$};
      \foreach \y in {1,2,3,4,5,6,7}
      {
        \draw (.15,\y) -- (-.15,\y) node [left] {$\y$};
      }
      \draw (-.15,8) node [left] {$x_2$};
      \draw (-.15,-.15) node [left,below] {$0$};  
          \filldraw (2,2) circle (4pt) node [below] {offset}; 
        \draw (5,4) node {$Z$};
\end{tikzpicture}
\end{center}
\caption{A priced zone $(Z,\zeta)$.  It is represented as a standard DBM for $Z$,
  plus the offset cost $+4$ (i.e. cost at the lowest point of the
  zone, here point $(2,2)$), plus the rate ($-1$) for $x$ and the rate
  for $y$ ($+2$). 
\label{fig:priced-zone}}
\end{figure}

Decidability and complexity for this model were proven in the early
papers~\cite{ATP01,BFH+01,BBBR07}, making use of a refinement of the
region construction (called corner-point abstraction later
in~\cite{BBL08}). Already in~\cite{LBB+01}, a symbolic solution built
on zones has been proposed, based on so-called \emph{priced
  zones}. The idea is to store information on the cost function on top
of a zone as a pair $(Z,\zeta)$, where $\zeta$ is an affine function
of the clocks, the meaning being that valuation $v \in Z$ can be
reached from the initial configuration, and the minimal cost to do
that is given by $\zeta(v)$. Similar to zones, priced zones can be
efficiently represented by a DBM, an offset cost and an affine
coefficient for each clock, see Figure~\ref{fig:priced-zone}.  All
operations needed for a forward exploration can be done using this
data structure. Only termination of the computation has remained open
until~\cite{BCM16}. Indeed, there is \textit{a priori} no possible
sound extrapolation in this context, but the development of the
simulation approach (as presented in Section~\ref{sec:simulation}) for
timed automata proved extremely useful and could be extended in some
sense to weighted timed automata as well~\cite{BCM16}.

\subsection{Liveness properties}

Liveness properties amount to checking whether some good event may
happen infinitely often, and are captured using B\"uchi conditions:
given a timed automaton $\A$ and a state $q$, does there exist an
infinite run of $\A$ that visits $q$ infinitely often? To solve this
problem using zones, one crucial modification needs to be done in both
the extrapolation and the simulation approaches. In the
\textsf{extra-Trans} rule, the inclusion $Z' \subseteq Z''$ is
replaced with $Z' = Z''$~\cite{Laarman:CAV:2013,Li:FORMATS:2009}. In
the $\preceq$-\textsf{Trans} rule, in addition to
$(q', Z') \preceq (q', Z'')$, we add $(q', Z'') \preceq (q', Z')$
which is a simulation in the other direction as
well~\cite{HSTW16,HSTW20} (simulation is replaced with a
bisimulation).

This seemingly simple change results in a huge difference in
performance.  It has been noticed in experiments that the number of
zones enumerated for liveness properties is substantially higher than
for reachability properties.  This has led to a close study of the
role of inclusion/simulation (collectively called subsumptions), as
opposed to equality/bisimulation, in pruning zones.  Suppose we call
the transition system over symbolic states computed using subsumptions
as a subsumption graph. Reachability problems can be solved in
polynomial time if the subsumption graph is given as input.  However,
it has been shown that deciding liveness is PSPACE-complete even if
both the automaton and a subsumption graph are given as
input~\cite{HSTW20}. This is evidence to the power of subsumptions in
reducing the number of symbolic states. Unfortunately, we cannot use
the full power of subsumptions for liveness.

It is however possible to restrict subsumption in such a way that no
spurious B\"uchi run is created.  Algorithms that can use restricted
subsumption for the liveness problem have been
studied~\cite{Laarman:CAV:2013,HSTW16,HSTW20}, with good performances
in practice.


\section{Tools}
\label{sec:tools}

Since the nineties, several tools have implemented 
efficient 
algorithms for the verification of timed automata, in particular:
\kronos{}~\cite{BDM+98}, \uppaal{}~\cite{BDL+06} and
\redlib{}~\cite{farn06} to cite a few.
In the last years we have started the development of a new open-source
tool \tchecker{}~\cite{TChecker}. In this section, we shortly present
the tools \uppaal{} and \tchecker{} which are both based on the zone
approach presented in the paper.

\subsection{A tool for verifying timed automata: \uppaal{}}

\uppaal{} is the state-of-the-art tool for timed automata
verification. Models consist of a network of timed automata that
communicate through handshaking and shared variables.
Specifications are expressed in a subset of the TCTL logic that allows
to express reachability properties as well as a restricted subset of
liveness properties.  Adding extra processes to the system allows to
check richer specifications.
The tool \uppaal{} implements model-checking algorithms. Extensions of
the tool have been proposed for statistical
model-checking~\cite{DBLP:journals/corr/abs-1207-1272} and two player
concurrent safety and reachability games~\cite{BCD+07}.

\uppaal{} has played a tremendous role in the adoption of the timed
automata formalism in the industry. Numerous case studies have been
successfully achieved using \uppaal{}: Philips Audio
Protocol~\cite{DBLP:conf/hybrid/LarsenPY95}, Bang\&Olufsen Audio/Video
Protocol~\cite{HSLL97}, Commercial Fieldbus
Protocol~\cite{DBLP:conf/ecrts/DavidY00}, Schedulability
analysis~\cite{DBLP:conf/isola/MikucionisLRNSPPH10}, Web Services
Business Activity~\cite{DBLP:conf/tacas/RavnSV11}, and SCADA Attacks
Detection~\cite{DBLP:conf/wetice/MercaldoMS19} to cite a few.
Several of these case studies have allowed to detect flaws in the
systems under study, leading to safer real-time systems.

Both the DBM library and the parser of \uppaal{} are
open-source. However, the remaining parts of its source code are not
publicly available.

\subsection{The tool \tchecker{}}

In the last years we have started the development of a fully
open-source verification tool for timed
automata. \tchecker{}~\cite{TChecker} consists of a set of libraries
and tools. It can be used both as a model-checker, and as a framework
to develop new verification algorithms.

\tchecker{} models consist in networks of timed automata that
communicate through multiprocess synchronisations and shared
variables. Its companion tool
\uppaaltotchecker{}~\cite{UppaalToTChecker} can be used to
automatically translate a subset of \uppaal{} input language into
\tchecker{} models.  Properties are encoded by adding observers to the
model. Reachability and liveness algorithms can be applied to detect
if specific states of the observers can be reached (repeatedly).

The \tchecker{} library is implemented in \texttt{C++} and provides
various classes that allow to manipulate models, to represent and
manipulate zones using DBMs, to compute zone graphs using various
semantics and extrapolations, and to represent and compute the
state-space of a model using verification algorithms.  The \tchecker{}
library and input language can easily be extended to support new data
structures and verification algorithms, as well as extensions of timed
automata such as weighted timed automata or timed games.  A tutorial
on how to implement verification algorithms using \tchecker{}
libraries is available from the website~\cite{TChecker}.

\tchecker{} also comes with a set of tools to perform syntactic
verification of models (\texttt{tck-syntax}), simulation
(\texttt{tck-simulate}), reachability verification
(\texttt{tck-reach}) and liveness verification
(\texttt{tck-liveness}). These tools can be used online on \tchecker{}
demonstration webpage~\cite{TCheckerOnline}.
The \tchecker{} libraries and tools implement most of the approaches
described in this paper, although some of them are not implemented yet
and will be available in future releases of \tchecker{}.  Several
research teams are currently using \tchecker{} to implement and test
their verification algorithms: Irisa (Rennes, France), LaBRI
(Bordeaux, France), LIF (Marseille, France) and CMI (Chennai, India).


\section{What next?}
\label{sec:next}

Over the years, timed automata verification has been successfully used
in many case studies,
e.g.~\cite{LPY95,HSLL97,DBLP:conf/ecrts/DavidY00,DBLP:conf/wetice/MercaldoMS19}.
In several instances complex bugs have been discovered and fixed,
leading to safer real-time systems. Efficient verification algorithms
and tools are crucial for practical applications. Timed automata
verification algorithms heavily rely on finite abstractions for
termination. As we have discussed in this article, this has been
achieved by means of extrapolations or simulations.  Both
extrapolations and simulations serve to keep the number of zones $Z$
per control state $q$ firstly finite, and secondly as small as
possible. This is the motivation behind finding coarser
extrapolations/simulations, and this endeavour has been largely
successful. Based on our experience with experiments, we find that by
applying extrapolation/simulation the number of zones per control
state is usually small. The major bottleneck is the large number of
control states $q$ appearing in the forward computation. In untimed
systems, this explosion is handled using SAT-based methods or
partial-order reduction. Lifting these techniques to the timed setting
has not been straightforward. We conclude this article by first
discussing some attempts in this direction and then finish with some
other open challenges.

\subsection{BDD and SAT based methods }
\label{sec:alternative}
In \cite{RSM-cav19}, abstract reachability algorithms were given where
zones were overapproximated by a set of clock constraint
predicates. This generalizes the zone-based exploration algorithm but
requires a counter-example guided refinement loop to ensure soundness
and completeness.  A similar idea was considered in
\cite{TM-formats17}, which is based on using lazy abstractions when
exploring the zones, and refining them using interpolants to ensure
that the exploration is sound and complete.

\cite{RSM-cav19} also gives a variant of the algorithm based on binary decision diagrams (BDDs), where the abstract zone-based semantics
was encoded using Boolean functions. 
A similar but more efficient way of using predicate abstractions for analyzing timed automata was given in \cite{CGMRT-cav19} based on
the use of the IC3 algorithm \cite{Bradley-vmcai11} combined with implicit abstractions \cite{CGMT-tacas14}.
Thanks to the implicit abstraction method, one does not need to build the abstract transition relation, but rather guesses, at each step,
an abstract transition and its corresponding witness concrete transition. This enables the use of SAT-based algorithms, including IC3.
The nuXmv model checker was extended with a syntax to describe timed automata, and provides a powerful alternative to zone-based algorithms
for timed automata models with large discrete state spaces.

There have been previous propositions for applying IC3 to timed automata without abstractions, such as \cite{KJN-formats12} where zones were used
to backpropagate counterexamples to inductions.

There have been other attempts at combining timed automata semantics with symbolic approaches for handling large discrete state spaces.
An algorithm encoding the region automaton using BDDs was given in \cite{KJN-acsd11} along with an extension of the SMV language for specifying timed automata.
In the past, several works attempted at extending binary decision diagrams to represent information on the clocks
such as \cite{Wang-ftnds01,beyer2003rabbit,EFGP-rtss2010}, while some of these only consider discrete time
\cite{NSLDL-fm12,thierry2015symbolic}.
In \cite{DDDHPSWW-scp12}, an extension of the \emph{and-inverter graph} data structure with predicates was considered in order to
represent the state space of linear hybrid automata. These can represent possible non-convex polyhedra extended by Boolean variables.

Bounded model checking was applied to timed automata in \cite{Sorea-entcs03,ACKS-forte02} and \cite{KJN-acsd11}; see also \cite{MN-tacas10} specifically for the partial-order semantics.

The above works conclude on the complementarity of the zone-based
enumerative algorithms and various symbolic approaches either based on
(variants of) BDDs or SAT/SMT solvers. This is, in fact, also observed
for finite-state models where enumerative and symbolic model checkers
are both useful in different contexts.  We thus believe that it is
important that both zone-based model checkers such as \uppaal and
\tchecker, and SAT-based ones such as \nuXmv are available.  A
practical verification tool box should in fact contain various
algorithms since a single algorithm may not succeed in verifying all
types of models.

\subsection{The local-time semantics and partial-order reduction}

Verification of very large networks with multiple timed processes is currently
out of reach of the existing methods. 
Enumerative model-checkers for untimed systems have immensely
benefitted from partial-order reduction methods that exploit the
concurrency in the network representation of the model. Roughly, two
actions $a$ and $b$ are independent if there is no component of the
network containing both these actions. For such actions, doing $ab$ or
$ba$ from a global state leads to the same global state (popularly
known as the diamond property). This property allows to pick one of
either $ab$ or $ba$ for further exploration. In general, given a set
of independent actions, partial-order methods aim to pick one path out
of the several possible interleavings of the independent actions. This
leads to exponential reduction in the number of states
enumerated. Unfortunately, for (networks of) timed automata,
partial-order reduction is not straightforward due to the lack of a
diamond property: for instance, suppose that $x$ is reset at $a$ and
$y$ is reset at $b$, then $ab$ and $ba$ lead to different zones
keeping track of the different order of resets. There is an implicit
synchronization due to the global nature of time.

In the last few years, there has been some work in making use of a
local-time semantics for networks of timed
automata~\cite{Bengtsson-local-time} to obtain a symbolic computation
that contains diamonds. The nice aspect of the local-time semantics is
the presence of diamonds, but unfortunately, getting finite
abstractions is not immediate. A first solution has been proposed
in~\cite{Govind:CONCUR:19}, but it is not compatible with
partial-order reduction. It has later been shown that there is no
finite abstraction for the local-time semantics~\cite{Govind:LICS:22}
that is compatible with partial-order reduction, although some
subclasses of timed automata admit such a finite abstraction. The next
challenge is to define a partial-order reduction technique which works
well in the timed setting, as well as to detect subclasses of timed
automata for which partial-order reduction and finite abstractions can
be combined.  Other works on partial-order reduction which do not go
via the local-time semantics also work on restricted settings:
applying the reduction only to parts of the network where independent
actions occur in zero time, \cite{DBLP:conf/atva/LarsenMMS20}, or
discovering independent actions in the standard global-time semantics,
either statically~\cite{DBLP:journals/fac/DamsGKK98} or
dynamically~\cite{DBLP:conf/cav/HansenLLN014}.

\subsection{Domain-specific algorithms} 
For some applications that can be modelled as timed automata, standard
algorithms for generic timed automata might be quite slow.  We
therefore believe that it would be beneficial to develop
domain-specific verification algorithms. As an example, the model of
funnel automata has been developed to model some robotics
systems~\cite{BMPS15,BMPS17}: funnel automata are timed automata with
few clocks (three clocks) but a large discrete state-space. While
small instances could be verified using tool \tiamo~\cite{BCM16} (a
precursor of \tchecker implemented in OCaml), larger instances could
not be verified due to the state-explosion problem. Techniques mixing
BDDs and zones could possibly be developed for this specific
application.

\subsection{Richer models}

Another interesting direction is to investigate the extent to which
the advances in timed automata can help in the algorithmics for richer
models. There has been some progress already in this direction. Apart
from updatable timed automata and weighted timed automata, the
simulation approach has been extended to pushdown timed
automata~\cite{AGP21} and event-clock automata \cite{AGGS22}. It
remains to be seen whether this approach can be lifted to the context
of parametric timed automata~\cite{PTA,IMITATOR}, probabilistic timed
automata~\cite{ProbTA,PRISM} and controller synthesis for timed
games~\cite{TimedGames,TIGA}.

\bibliographystyle{splncs04}
\bibliography{formats22}

\end{document}